\documentclass{PoS}

\usepackage{amsmath}
\usepackage{amssymb}

\newcommand{\bea}{\begin{eqnarray}}
\newcommand{\eea}{\end{eqnarray}}

\newcommand{\1}{\mbox{1}\hspace{-0.25em}\mbox{l}}

\newcommand{\scalar}{\bar{\psi}\psi}
\newcommand{\dt}{\Delta t}
\newcommand{\dts}{{\Delta t}_s}
\newcommand{\NssN}{\left\langle N|\bar ss|N\right\rangle}

\title{   \begin{picture}(0,0)(0,0)%
   \put(355,75){\makebox(0,0)[l]{\textnormal{\normalsize UTHEP-620}}}%
   \end{picture}%  
 Nucleon strange quark content in $2+1$-flavor QCD}

\ShortTitle{Nucleon strange quark content in $2+1$-flavor QCD}

\author{
JLQCD collaboration:\,\, 
\speaker{K. Takeda}$^a$, 
S.~Aoki$^{a,b}$, 
S.~Hashimoto$^{c,d}$, 
T.~Kaneko$^{c,d}$,
T.~Onogi$^e$
and 
N.~Yamada$^{c,d}$ 
\\%\thanks{A footnote may follow.}\\
\llap{$^a$}Graduate School of Pure and Applied Sciences,
           University of Tsukuba, Tsukuba, Ibaraki 305-8571, Japan\\
\llap{$^b$}Center for computational Sciences, University of Tsukuba, Tsukuba, 
  Ibaraki 305-8577, Japan
\llap{$^c$}High Energy Accelerator Research Organization (KEK), 
           Tsukuba 305-0801, Japan\\
\llap{$^d$}School of High Energy Accelerator Science, 
           The Graduate University for Advanced Studies\\
           (Sokendai), Tsukuba 305-0801, Japan\\
\llap{$^e$}Department of Physics, Osaka University Toyonaka, 
           Osaka 560-0043, Japan\\
E-mail: \email{ktakeda@het.ph.tsukuba.ac.jp}
}

\abstract{
We calculate the strange quark content of the nucleon $\NssN$ 
directly from its disconnected three-point function in $N_f\!=\!2+1$ QCD.
Chiral symmetry is crucial to avoid a possibly large contamination
due to operator mixing, 
and is exactly preserved by employing the overlap quark action.
We also use the all-to-all quark propagator and 
the low-mode averaging technique
in order to accurately calculate the relevant nucleon correlator. 
Our preliminary result extrapolated to the physical point 
is $f_{T_s} \!=\! m_s \NssN/ M_N = 0.013(12)(16)$,
where $m_s$ and $M_N$ are the masses of strange quark and nucleon.
This is in good agreement with our previous estimate in $N_f\!=\!2$ QCD
as well as those from our indirect calculations 
using the Feynman-Hellmann theorem.
}

\FullConference{
The XXVIII International Symposium on Lattice Field Theory, Lattice2010\\
June 14-19, 2010\\
Villasimius, Italy
}

\begin{document}

\section{Introduction}

The strange quark content of the nucleon $\NssN$ is a fundamental
quantity on the nucleon structure. It represents
the effect of strange quark on the nucleon mass $M_N$ 
\bea
   f_{T_s} = \frac{m_s \NssN}{M_N},
\eea
where $m_s$ is the strange quark mass.
This parameter is also relevant to direct experimental searches
for the dark matter, as one of its
candidates, namely neutralino,  
may interact with the nucleon most strongly 
through the strange quark content.
Therefore,
$\NssN$ is crucial 
to assess the sensitivity of the experiments~\cite{Ellis:2008hf}.

We calculate $\NssN$ from lattice QCD simulations in this study.
The relevant nucleon three-point function is purely composed of 
a disconnected diagram, 
whose computational cost is prohibitively high
with the conventional simulation methods.
We overcome this difficulty by using 
the methods of the low-mode averaging ~\cite{DeGrand:2004qw,Giusti:2004yp}
and the all-to-all propagator~\cite{Foley:2005ac}.

Another advantage of our study is that
chiral symmetry is exactly preserved by using the overlap quark action.
We point out in this article that 
explicit symmetry breaking with conventional lattice fermions
induces operator mixing 
leading to a possibly large contamination in $\NssN$. 

\section{Simulation setup}
\label{sec:sim}

We simulate $N_f\!=\!2+1$ QCD 
with the Iwasaki gauge and the overlap quark actions.
The lattice spacing is determined as $a = 0.112(1)$~fm 
using the $\Omega$ baryon mass as an input.
We simulate two values of degenerate up and down quark masses 
$m_{ud}\!=\!0.035$ and 0.050
on a $N_s^3 \!\times\! N_t = 16^3 \times 48$ lattice.
The strange quark mass is set to $m_s=0.080$ and $0.100$,
which are close to the physical mass $m_{s,phys}\!=\!0.081$
fixed from $M_K$. 
We also push our simulations to lighter masses $m_{ud}\!=\!0.015$ and 0.025
on a larger lattice $24^3 \times 48$ 
but with a single value of $m_s\!=\!0.080$. 
The four values of $m_{ud}$ cover a range of $M_\pi\sim 290-520$~MeV
with a condition $M_\pi L \gtrsim 4$ satisfied.
We have accumulated about 50 independent gauge configurations 
at each combination of $m_{ud}$ and $m_s$.

We calculate the nucleon three-point function
\begin{eqnarray}
   C_{3pt}^\Gamma
    ({\bf y},t_{src},\dt,\dts)
   & = &
   \frac{1}{N_s^6} 
   \sum_{\bf x,z}
   \left\{
      \mathrm{tr}_s \left[         \Gamma 
    \langle 
             N({\bf x},t_{src}+\dt)\bar{s}s({\bf z}, t_{src}+\dts)
	     \overline{N}({\bf y},t_{src})
	     \rangle
      \right]
   \right.
\notag \\
   && 
   \left.
      \hspace{7mm}
     - \langle \bar{s}s({\bf z},t_{src}+\dts) \rangle \,
        \mathrm{tr}_s\left[          \Gamma 
     \langle 
           N({\bf x},t_{src}+\dt) \overline{N}({\bf y},t_{src})
         \rangle
      \right]
   \right\},  
\label{3pt}
\end{eqnarray}
where $N=\epsilon^{abc}(u_a^TC\gamma_5d_b)u_c$ is 
the nucleon interpolating operator, 
and $\bar{s}s$ is the strange scalar operator on the lattice.
We denote the temporal separation between the nucleon source 
and sink (scalar operator) by $\dt$ ($\dts$).
This correlator is calculated with four choices of the source location, 
namely $t_{src}=0,12,24$ and $36$ with ${\bf y}=0$,
and two choices of the projection operator
$\Gamma=\Gamma_\pm=(1\pm\gamma_4)/2$,
which correspond to the forward and backward propagating nucleons. 
We then take the average 
\bea
   C_{3pt}(\dt,\dts) 
   & = & 
   \frac{1}{8} 
   \sum_{t_{src}} 
   \left\{
      C_{3pt}^{\Gamma_+}({\bf y},t_{src},\dt,\dts) 
     +C_{3pt}^{\Gamma_-}({\bf y},N_t-t_{src},N_t-\dt,N_t-\dts) 
   \right\}
\eea
to reduce its statistical fluctuation.
We also calculate the nucleon two-point function $C_{2pt}(\dt)$ 
in a similar way.

%%% All-to-All %%%%%%%%%%%%%%%%%%%%%%%%%%%%%%%%%%%%%%%%%%%%%%%%%%%%%%%

We calculate the disconnected strange quark loop from $\bar{s}s$
and its vacuum expectation value $\langle \bar{s}s\rangle$
using the all-to-all propagator.
It is expected that 
low-lying modes of the Dirac operator $D$ contribute 
dominantly to low-energy dynamics of QCD,
and we calculate this contribution exactly 
\begin{equation}
   (D^{-1})_{low}(x,y)
   =
   \sum_{i=1}^{N_e}\frac{1}{\lambda^{(i)}}v^{(i)}(x)v^{(i)}(y)^{ \dagger}
   \hspace{5mm} 
   (D v^{(i)} = \lambda^{(i)} v^{(i)}),
   \label{DovL}
\end{equation}
where the number of eigenmodes $N_e$ is 160 (240) 
on our smaller (larger) lattice.
The contribution of the remaining high-modes is estimated stochastically.
We generate a single complex $Z_2$ noise vector $\eta(x)$ 
for each configuration,
and divide it into $N_d=3\times
4\times N_t/2$ vectors $\eta^{(d)}(x)$, which have non-zero elements
only for a single combination of color and spinor indices on two
consecutive time-slices.
The high-mode contribution is then estimated as 
\begin{equation}
 (D^{-1})_{high}(x,y)
  =
  \sum_{d=1}^{N_{d}}\psi^{(d)}(x)\eta^{(d)}(y)^{\dagger},
  \label{DovH}
\end{equation}
where $\psi^{(d)}(x)$ is obtained by solving 
\begin{equation}
   D \psi^{(d)}(x) = (1-\mathcal P_{low})\eta^{(d)}(x), 
   \hspace{5mm}
   \mathcal P_{low}=
   \sum_{i=1}^{N_e}v^{(i)}(x)v^{(i)}(y)^{ \dagger}.
\end{equation} 
We note that, in this report, 
$C_{3pt}$ on $24^3 \times 48$ is calculated without the high-mode contribution
in the quark loop,
our measurement of which is currently underway.
In our study in $N_f\!=\!2$ QCD \cite{Takeda:2009,Takeda:2010},
we observed that 
the result for $\NssN$ is well dominated by the low-modes
and does not change significantly by ignoring the high-mode contribution.

%// LMA %%%%%%%%%%%%%%%%%%%%%%%%%%%%%%%%%%%%%%%%%%%%%%%%%%%%%%%%%%%%%%%

We can improve the statistical accuracy of the nucleon correlators by using LMA.
Let us divide $C_{2pt}$ 
into two contributions $C_{2pt,low}$ and $C_{2pt,high}$: 
$C_{2pt,low}$ is the two-point function constructed only from 
the low-mode part of the quark propagator $(D^{-1})_{low}$,
and $C_{2pt,high}$ is the remaining contribution.
We can average over all possible source points $({\bf y},t_{src})$ for
$C_{2pt,low}$.  
The piece representing the nucleon propagation in $C_{3pt}$ 
can be calculated in a similar way.
In this study, we average over all $t$ for $t_{src}$ but 16 spatial locations
for ${\bf y}$ at each $t_{src}$.

%%% smearing %%%%%%%%%%%%%%%%%%%%%%%%%%%%%%%%%%%%%%%%%%%%%%%%%%%%%%%%%

In our previous study in $N_f\!=\!2$ QCD,
we observed that the smearing of the nucleon source and sink operators $N$
is crucial to obtain a clear signal of $C_{3pt}$ 
at reasonably small $\dt$.
We therefore construct $N$ from the quark field with the Gaussian smearing
\begin{equation}
 q_{smr}^{}({\bf x},t) 
  =
  \sum_{\bf y} \left\{ 
		\left( {\1}+\frac{\omega}{4N}H \right)^N 
		\right\}_{{\bf x,y}} 
  q({\bf y},t), \label{eqn:Gaussian} \qquad
  H_{{\bf x,y}} 
  =
  \sum_{i=1}^3 (\delta_{{\bf x,y}-\hat i}+\delta_{{\bf x,y}+\hat i}),
\end{equation}
where the parameters are set to $\omega=20$ and $N=400$.
This smeared operator is not gauge invariant but we fix the gauge
to the Coulomb gauge. We use $q_{smr}$ for both $C_{3pt}$ and $C_{2pt}$.

\section{Strange quark content at simulated quark masses}

\label{Sec:Extraction}
\begin{figure}[tbp]
  \centering
  \includegraphics[width=0.46\textwidth,clip]{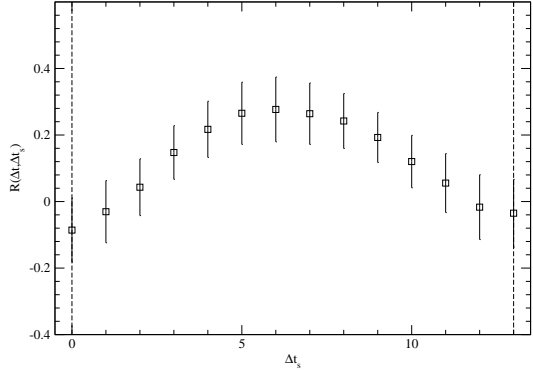}
  %\vspace{-3mm}
  \caption{
    Ratio $R(\dt=13,\dts)$  
    at $m_{ud}=0.050$ and $m_{s}=0.080$
    as a function of $\dts$. 
    The vertical lines show the locations of the nucleon operators.
    The noisy high-mode contribution to the quark loop is omitted in
    this plot. 
  }
  \label{Fig:smr_m050_m080_dt13_l}
\end{figure}

\begin{figure}[tbp]
  \centering
  \includegraphics[width=0.46\textwidth,clip]{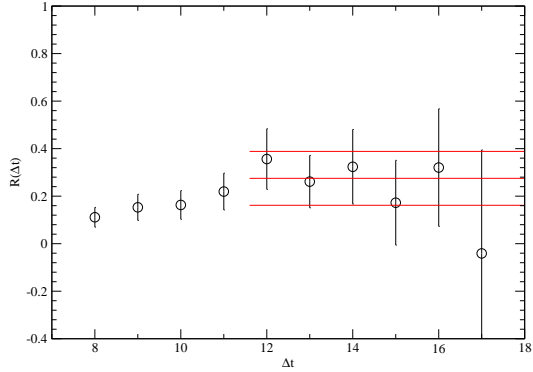}
  %\vspace{-3mm}
  \caption{
    Fit result $R(\dt)$ at $m_{ud}=0.050$ and $m_{s}=0.080$
    as a function of $\dt$.
    The horizontal lines show the result of a constant fit in $\dt$
    and its error band.
%    We also plot $R(\dt)$ at $\dt \leq 10$ obtained by a constant fit 
%    to $R(\dt,\dts)$ in $\dts \in [???,???]$.
  }
  \label{Fig:fit_dt_m050_m080}
\end{figure}

We extract the strange quark content $\NssN$ on the lattice 
from the ratio of $C_{3pt}$ and $C_{2pt}$
\begin{equation}
    R(\dt,\dts) 
   \equiv \frac{C_{3pt}(\dt,\dts)}{C_{2pt}(\dt)}
   \xrightarrow[\dt, \dts \to \infty]{} 
   \langle N|\bar{s}s|N\rangle.
   \label{Eqn:ratio}
\end{equation}
In order to identify a plateau of $R(\dt,\dts)$,
it is helpful to consider the same ratio but 
approximated by taking only the low-mode contribution in the scalar loop.
Figure~\ref{Fig:smr_m050_m080_dt13_l} shows 
an example of the approximated ratio as a function of $\dts$. 
The significant change near $\dts \sim 0$ and $\dt$ 
is due to a contamination from excited states.
We therefore fit $R(\dt,\dts)$ to a constant form 
in $\dts \!=\! [5,\Delta t-5]$, 
which is well separated from the nucleon source and sink.
Note that this fit and the following analysis on $16^3 \times 48$
are carried out using $R(\dt,\dts)$ without the approximation.
While we use the approximated ratio on $24^3 \times 48$ 
as mentioned in the previous section,
we confirm in the analysis on $16^3 \times 48$
that the high-mode contribution to the quark loop
has only small effect to $\NssN$.

The result of the constant fit in terms of $\dts$ is denoted as $R(\dt)$
and is plotted as a function of $\dt$ in Fig.~\ref{Fig:fit_dt_m050_m080}.
The stability of $R(\dt)$ at $\dt \geq 12$ suggests that 
these data are well dominated by the ground state contribution. 
We extract $\NssN$ by a constant fit to $R(\dt)$ in this region.

\section{Chiral extrapolation to the physical point}

\begin{figure}[tbp]
  \centering
  \includegraphics[width=0.46\textwidth,clip]{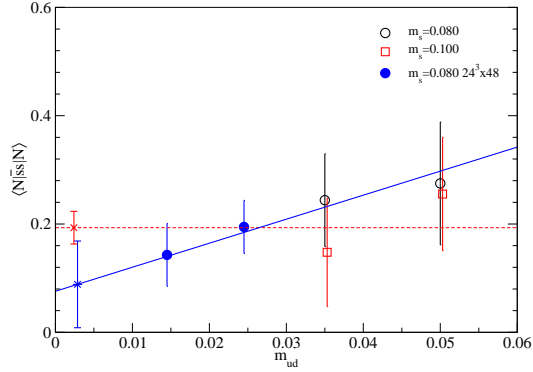}
  %\vspace{-3mm}
  \caption{
   Strange quark content $\NssN$ as a function of $m_{ud}$.
   Circles and squares show data at $m_s\!=\!0.080$ and 0.100,
   whereas filled and open symbols are obtained on 
   $16^3 \times 48$ and $24^3 \times 48$, respectively.
   We also plot the linear and constant fits by solid and dashed lines.
  }
  \label{Fig:extrap_nf3.lnr}
\end{figure}

In Fig.~\ref{Fig:extrap_nf3.lnr}, 
we plot the matrix element $\NssN$ as a function of $m_{ud}$.
Our data show a mild dependence on both $m_{ud}$ and $m_s$,
which can be well described by a linear form 
\bea
   \langle N|\bar{s}s|N\rangle
   = c_0+c_{1,ud}m_{ud}+c_{1,s}m_{s}
   \label{eqn:linear}
\eea
with $\chi^2/{\rm d.o.f.}\!\sim\!0.06$.
While we also attempt a chiral extrapolation based on heavy baryon 
chiral perturbation theory \cite{WalkerLoud:2004hf}, 
the chiral expansion turns out to have a poor convergence 
up to the next-to-next-to-leading order at our simulated quark masses.
We therefore use the linear chiral extrapolation.
The systematic error due to the choice of the fitting form
is estimated by comparing with a constant fit,
which also leads to a reasonable value of $\chi^2/{\rm d.o.f.}\!\sim\!0.44$.
We obtain $\NssN\!=\!0.086(81)(107)$ at the physical point,
where the first and second errors are statistical and systematic,
respectively.

This bare matrix element is converted to the renormalization invariant
parameter
\begin{eqnarray}
  f_{T_s}
  =
  \frac{m_s \NssN}{M_N}
  =
  0.013(12)(16).
  \label{Eqn:fts}
\end{eqnarray}
We note that this is in good agreement with our previous estimate
in $N_f\!=\!2$ QCD \cite{Takeda:2009,Takeda:2010}
as well as our indirect determinations 
through the Feynman-Hellmann theorem 
in $N_f\!=\!2$ \cite{Ohki:2008ff} and $N_f\!=\!2+1$ QCD \cite{Ohki:2009}.

\section{Renormalization of $\bar{s}s$}

In (\ref{Eqn:fts}), 
$f_{T_s}$ is expressed with the bare quantities
assuming that the operator $m_s\bar{s}s$ is renormalization invariant
as in the continuum limit.
This is, however, not the case  
if chiral symmetry is explicitly broken by the lattice fermions of the choice.
To illustrate this,
let us consider the renormalization of the scalar operator $\bar{s}s$ 
in the flavor $SU(3)$ symmetric limit for simplicity.
Using the flavor triplet quark field $\psi$, 
$\bar{s}s$ can be written as a linear combination of 
the flavor singlet and octet scalar operators
\begin{equation}
  (\bar{s}s)^{phys} 
  = 
  \frac{1}{3}
  \left\{
   (\scalar)^{phys} 
   -\sqrt{3}\, (\bar{\psi}\lambda^8 \psi)^{phys}
  \right\},
\label{eqn:singlet_octet}
\end{equation}
where $\lambda^8$ is a Gell-Mann matrix,
and we put the superscript ``$phys$'' to the renormalized operators 
to distinguish them from bare lattice operators.
The singlet and octet operators are renormalized as 
\begin{eqnarray}
 (\scalar)^{phys}
  &=& 
  Z_0\, (\scalar)
  \label{singlet},\\
  ( \bar{\psi}\lambda^8 \psi)^{phys}
  &=& 
  Z_8\, ( \bar{\psi}\lambda^8 \psi),
  \label{octet}
\end{eqnarray}
in renormalization schemes which respect chiral symmetry.
The operator $(\bar{s}s)^{phys}$ is thus expressed as
\begin{equation}
   (\bar{s}s)^{phys}  =
   \frac{1}{3}
   \left\{
      (Z_0 + 2 Z_8) (\bar{s}s)
     +(Z_0 - Z_8)(\bar{u}u+\bar{d}d)
   \right\}.
   \label{eqn:sbars:gen}
\end{equation}
This implies that $\bar{s}s$ can mix with the up and down quark
operators unless $Z_0=Z_8$. 

\begin{figure}[tbp]
  \centering
  \includegraphics[width=0.36\textwidth,clip]{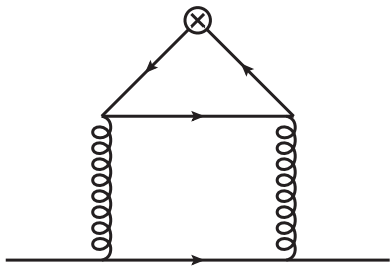}
  %\vspace{-3mm}
  \caption{
  Disconnected diagram contributing the renormalization of the flavor
  singlet scalar operator (cross). In higher orders, 
  more gluons propagate between the quark loop and propagator on the bottom.
 }
  \label{Fig:diagram}
\end{figure}

The difference $Z_0\!-\!Z_8$ comes from disconnected diagrams,
such as that in Fig.~\ref{Fig:diagram},
contributing only to $Z_0$.
These diagrams are in fact forbidden by chiral symmetry: 
the quark loop with the scalar operator 
$\bar{s}s=\bar{s}_R s_L+\bar{s}_L s_R$ has to flip the chirality,
while such a flip does not occur 
with the quark-quark-gluon vertices in the diagrams.
This implies $Z_0\!=\!Z_8$,
which also holds at finite quark masses 
in mass independent renormalization schemes.
The renormalization of $\bar{s}s$ thus reduces to
a simple multiplicative renormalization 
\bea
   (\bar{s}s)^{phys} = Z_S \bar{s}s, 
   \hspace{5mm}
   Z_S\!=\!Z_8\!=\!Z_0,
   \label{eqn:renorm}
\eea
and the operator $m_s \bar{s}s$ is renormalization invariant 
provided that 
chiral symmetry is exactly preserved as in our study.

When chiral symmetry is explicitly broken by a lattice fermion formulation,
(\ref{eqn:renorm}) is modified as 
\begin{equation}
   (\bar{s}s)^{phys} 
   = 
   \frac{1}{3}
   \left[
      (Z_0 + 2 Z_8) (\bar{s}s)^{lat} 
      +(Z_0 - Z_8)(\bar{u}u+\bar{d}d)^{lat}
     + \frac{b_0}{a^3} + \cdots
   \right].
   \label{eqn:renom:Wilson}
\end{equation}
The second term represents the mixing with the light quark contents 
due to $Z_0\!\ne\!Z_8$. This may lead to a large contamination 
in $(\bar{s}s)^{phys}$, because the light quark contents 
induce a connected diagram,
the magnitude of which is much larger than that of the disconnected one.
Previous direct calculations with the Wilson-type fermions 
% \cite{Fukugita:1994,Dong:1996,Gusken:1998,Michael:2001} 
\cite{Fukugita:1994,Dong:1996,Gusken:1998} 
in fact obtained rather large values $f_{T_s}\!\sim\!0.3$\,--\,0.5 
compared to our result (\ref{Eqn:fts})
probably due to this contamination.

The explicit symmetry breaking also induces 
the mixing of the flavor-singlet operator $\scalar$
with lower dimensional operators, 
such as the third term in (\ref{eqn:renom:Wilson}),
through the renormalization (\ref{singlet}).
This contribution must be subtracted 
as a part of the vacuum expectation value of $\bar{s}s$.
Due to the cubic divergence, 
this results in a large cancellation,
with which the calculation is potentially noisier.

\section{Conclusion}
We calculate the nucleon strange quark content in $N_f\!=\!2+1$ QCD.
Chiral symmetry is exactly preserved by employing the overlap quarks
in order to avoid large contaminations from the operator mixing.
We also use the LMA and the all-to-all propagator to accurately calculate
the relevant nucleon correlators.
Our result of $f_{T_s}$ is in good agreement with 
our previous study in $N_f\!=\!2$ QCD as well as 
our indirect calculations for $N_f\!=\!2$ and 3 
using the Feynman-Hellmann theorem:
all our studies consistently favor 
small strange quark content $f_{T_s}\!\approx\!0.02$.

Numerical simulations are performed on Hitachi SR11000 and 
IBM System Blue Gene Solution 
at High Energy Accelerator Research Organization (KEK) 
under a support of its Large Scale Simulation Program (No.~09/10-09).
This work is supported in part by the Grant-in-Aid of the
Ministry of Education,  Culture, Sports,
Science and Technology, (No.~20340047, 21674002 and  21684013)
and by the  
Grant-in-Aid for Scientific Research on Innovative Areas, 
(No.~20105001, 20105002, 20105003 and 20105005).

\end{document}